\documentclass[11pt,preprint,flushrt]{aastex}
\usepackage{amsmath}
\usepackage{amssymb}
\usepackage{graphics}
\usepackage{natbib}

\shortauthors{Geroyannis}
\shorttitle{THE TURN-OVER SCENARIO FOR WHITE DWARFS AND NEUTRON STARS}

\begin{document}

\title{THE TURN-OVER SCENARIO FOR WHITE DWARFS \\
       AND NEUTRON STARS}

\author{V. S. Geroyannis}

\affil{Astronomy Laboratory, Department of Physics, University of Patras, Greece,}
\affil{GR-26500 PATRAS, GREECE}

\email{vgeroyan@physics.upatras.gr}

\begin{abstract}
We study numerically the so-called ``turn-over scenario'' for rotating magnetic white dwarfs and neutron stars. According to this scenario, the magnetic symmetry axis of the model inclines at a gradually increasing angle (the so-called ``turn-over angle'') relative to the invariant angular momentum axis. Consequently, the model becomes ``perpendicular rotator'' (i.e., its turn-over angle becomes almost $90\arcdeg$) on a ``turn-over timescale'' calculated to be $\sim \textrm{(a few)} \times 10^6 \ \textrm{--} \ \textrm{(a few)} \times 10^7 \textrm{~yr}$ for the examined white dwarf models and $\sim 10^1 \ \textrm{--} \ 10^3 \textrm{~yr}$ for the examined neutron star models. Furthermore, the initial differential rotation of the model turns to uniform rotation due to angular momentum mixing caused by hydrodynamic Alfven waves propagating along the poloidal magnetic field lines. Our numerical results show that, during the turn-over phase, the spin-down time rate is large, while the spin-down power remains small; so, the turn-over phase is a characteristic case for which an eventually observed large spin-down time rate should not be interpreted as implying a large spin-down power. 
\end{abstract}


\section{Introduction}
``Turn-over'' or ``turn-over phase'' or ``turn-over process''  is 
called the gradual increase of the angle between the magnetic symmetry
axis and the rotation axis of a star from $\sim 0\arcdeg$ (axisymmetric or aligned rotator) to $\sim90\arcdeg$ (orthogonal or perpendicular rotator).
Any intermediate configuration is called ``oblique rotator''.
The angle $\chi$ between the magnetic and the rotation axis is
called ``turn-over angle'' or ``obliquity angle''.
We say that we study a ``turn-over scenario'' if we take into
account the turn-over phase in this scenario.

Actually, during the turn-over phase, it is the angular
momentum that remains invariant (its axis is taken to be the $z$
axis); while the angular velocity is inclined at a small angle $\gamma$
(less than $2\arcdeg$, say) with respect to the angular momentum
axis. At a first approximation and without loss of generality,
we can assume that these two axes coincide.

Turn-over seems to be a rather underestimated phenomenon. Indeed, searching the NASA Astrophysics Data System (\url{http://adswww.harvard.edu}), we find that, during the last 30 years, $\sim 12000$ papers have been published with the keywords \texttt{(white AND dwarf) OR (neutron AND star) OR (pulsar)} in their title; while, at the same time, roughly 100 papers have in their title the keywords \texttt{(oblique AND rotator) OR (perpendicular AND
rotator)}. This is strange enough, since turn-over is closely related to
the well-known lighthouse model: as the pulsar spins on its axis, the
emission beam (with symmetry axis coinciding with the magnetic symmetry
axis of the star) sweeps the space, crossing periodically our
line of sight. This happens only if  $\chi > 0$, since, otherwise
(that is, for an aligned rotator), we can observe just a
radiation source coinciding with one of the magnetic poles of
the emitting star. So, an aligned rotator lacks the most
significant characteristic of a typical pulsar.

In the turn-over scenario, the casting of the roles has mainly to do with rotation, toroidal field, poloidal field, and turbulent viscosity. In particular, \textbf{rotation} (1) builds up the most efficient energy reservoir, i.e., the rotational kinetic energy, which then powers on a variety of interesting phenomena; and (2) induces oblate configurations due to centrifugal forces.

The \textbf{poloidal magnetic field} (1) turns on the magnetic dipole mechanism, which then converts rotational kinetic energy into radiated energy; and (2) cooperates with rotation in inducing oblate configurations, i.e., both these partners tend to derive oblate configurations.

The \textbf{toroidal magnetic field} (1) stabilizes the poloidal magnetic field structure; and (2) tends to derive prolate configurations, that is, it has opposite action to both rotation \& poloidal field.

Finally, the \textbf{turbulent viscosity} causes drastic energy dissipation and, thus, brings the configuration from an unstable high energy level (aligned rotator) to a stable low energy level (perpendicular rotator).

\section{The turn-over scenario}
A full discussion on this subject can be found in \citet[hereafter Paper I; symbols and definitions used here are identical to those in Paper I]{ger01} and also in \citet[hereafter Paper II]{ger02}, \citet[hereafter Paper III]{gpa02}, and \citet[hereafter Paper IV]{gpv02}. A brief description of the turn-over scenario has as follows. \\
\leftline{ }
\textbf{Early evolutionary phase (EEP)} \\ 
\textbf{(1)} We assume an axisymmetric differentially rotating magnetic
model, undergoing an axisymmetric early evolutionary phase of
secular timescale, $t_{SEC}$, during which rotation \& poloidal field
dominate over the toroidal field and derive oblate
configurations with moment of inertia $I_{33}$ (where the principal axis $I_3$ coincides with the spin axis) greater than $I_{11} (=I_{22})$; the other two principal axes $I_1$ and $I_2$ lie on the equatorial plane. \\
\textbf{(2)} During the EEP, the model suffers from secular angular
momentum loss and spin-down due to the magnetic dipole
mechanism, activated by the poloidal field; this leads to
secular decrease of both ellipticity and $I_{33}$. \\
\leftline{ } 
\textbf{Late evolutionary phase (LEP)} \\
\textbf{(1)} At some particular time signaling the beginning of a late
evolutionary phase, the toroidal field starts prevailing over
rotation \& poloidal field, first succeeding in equating the
moments of inertia, $I_{11} = I_{33}$, and then in establishing
``dynamical asymmetry'', $I_{11} > I_{33}$. \\
\textbf{(2)} ``Dynamically asymmetric configurations'' (DAC) tend to turn over spontaneously, so that to rotate about axis with moment of
inertia greater than $I_{33}$, with the angular momentum remaining
invariant; thus, the fate of a DAC is to become oblique rotator
and, eventually, perpendicular rotator. \\
\leftline{ }
\textbf{Turn-over phase (TOP)} \\ 
\textbf{(1)} During the turn-over phase, the turn-over angle, $\chi$, increases spontaneously up to $\sim 90\arcdeg$ on a turn-over timescale, $t_{TOV}$. The terminal model rotates about its $I_1$ axis, coinciding with the invariant angular momentum axis, and occupies the state of least energy consistent with its angular momentum and magnetic field. \\
\textbf{(2)} The excess energy due to differential rotation, defined by
the angular velocity component $\Omega_3$ along the spontaneously
turning over $I_3$ axis, coinciding in turn with the magnetic symmetry axis, is dissipated down to zero due to the very efficient action of
turbulent viscosity in the convective zone of the model. So, the
terminal model does not rotate about its magnetic symmetry axis. \\
\textbf{(3)} It seems very difficult for the terminal model to sustain
differential rotation along its $I_1$ axis, mainly due to the
destructive action of the poloidal field. There is a competition
between the efforts of the magnetic stresses to remove
rotational nonuniformities, and those of the rotational
velocities to bury and destroy the magnetic flux. If the
magnetic field and the electrical conductivity have appropriate
values, then the magnetic field prevails and removes all the
nonuniformities of rotation. So, the terminal model rotates
rigidly about its $I_1$ principal axis with angular velocity $\Omega_1$.

A critical question is \textit{if the magnetic field is strong enough to remove differential rotation}. This interesting problem is discussed in full detail in Paper I (\S~7); the discussion reveals a critical issue concerning the starting model. In particular, it seems that the toroidal magnetic field must be strong enough so as to induce an adequate dynamical asymmetry for the turn-over to take place; while, at the same time, it must be weak enough so as to permit interchanging of angular momentum via hydromagnetic Alfv\'en waves propagating across its field lines. The difficulty is that, on the one hand, dynamical asymmetry increases with the toroidal magnetic field and, on the other hand, efficiency in interchanging angular momentum decreases as the toroidal field increases. In view of this remark and to the extent that our problem concerns mainly the behavior of the starting model, the right question seems to be: \textit{is the toroidal magnetic magnetic field strong enough to induce dynamical asymmetry and weak enough to permit angular momentum mixing?}.

An approximative estimate for the turn-over timescale, $t_{TOV}$, is given by \citet[hereafter MT72, eq. (54); see also some interesting remarks regarding differential rotation at the last two paragraphs of \S~3]{mta72}
\begin{equation} 
t_{TOV} \simeq 
  \frac{D\!E_{DRD}}{V}
  \left \langle \Omega_3 \right \rangle^{-2}
  \left \langle \frac{H^2}{8 \pi \varrho a^2} \right \rangle^{-2} 
  \left \langle \frac{1}{3} \, \frac{\varrho V_t}{\Lambda_t} \right \rangle_{sz}^{-1}
  \left \langle \frac{\Omega_3^2 r^2}{a^2} \right \rangle_{sz}^{-2}
  \left \langle \Lambda_t \right \rangle_{sz}^{-2} \, , 
\end{equation}
where the subscript $sz$ denotes that the corresponding average is calculated over the surface zone with base at the transition layer, $\xi_s$, and top at the boundary, $\xi_1$, of the star; averages without subscripts are calculated over the whole star, i.e., they are global averages. This estimate results as the mean energy per unit volume available for dissipation, $D\!E_{DRD} / V$ ($D\!E_{DRD}$ is the excess rotational kinetic energy due to differential rotation and $V$ is the volume of the configuration), divided by the mean energy dissipation  during the turn-over due to turbulent viscosity in the convective surface zone per unit volume and per unit time (i.e., the product of the last 5 terms in the right-hand side of eq. [1]; details on the derivation of such a relation are given in MT72, eqs. [7], [49]--[50]).  

\section{Some results}
Table 1 gives some significant quantities involved in the turn-over scenario. Our computations concern two models of compact stars. 

The first model simulates the AE Aqr white dwarf (for white dwarf models of several masses, angular momenta, and magnetic fields, see Paper II [especially \S~7, Table 1], Paper III [especially \S~3, Figs. 1--4, Tables 5, 10, 15], and Paper IV [\S~3, Table 1]); the corresponding computations have been performed by the so-called ``complex plane iterative technique'' (CIT) \citep[and references therein]{gpa00}. 

The second model simulates the 1987A neutron star (for another neutron star model, see Paper IV [\S~3, Table 1]); the corresponding computations should be considered as rough computations. An accurate numerical treatment of neutron stars is now in preparation.

The results for the AE Aqr white dwarf can simplify the study of the well-known spin-down problem in this white dwarf. In particular, during the turn-over phase, the observed spin-down time rate (interpreted as spin-down time rate due to turn-over) is large, $\dot{P} \simeq 5 \times 10^{-14} \, \mathrm{s \, s^{-1}}$; while the estimated spin-down power, $\dot{T} \simeq 5 \times 10^{32} \, \mathrm{erg \, s^{-1}}$, is almost two orders of magnitude less than the classically estimated spin-down power (for details, see Paper I, \S\S~1--2). So, an interesting conclusion is that: \textit{an observed large spin-down time rate does not always imply a large spin-down power}.

Our results for the 1987A neutron star show that it needs $t_{SEC} \simeq 8.5 \, \textrm{yr}$ to complete its early enolutionary phase, and $t_{now} \simeq 8.5 \, \textrm{yr}$ to obtain an observationally favourable turn-over angle (in accordance with the lighthouse model), $\chi \sim 60\arcdeg$; the full turn-over timescale for this star is $t_{TOV} \simeq 16 \, \textrm{yr}$. In conclusion, \textit{we expect that we will be able to observe the 1987A neutron star as typical pulsar within the period 2004 (i.e., $1987+8.5+8.5 \, \textrm{yr}$) -- 2012 (i.e., $1987+8.5+16 \, \textrm{yr}$)}. 

\begin{table}
\begin{center}
\caption{Summary of calculations for the AE Aqr white dwarf (accurate computations) and  the 1987A neutron star (rough computations). Parameters in cgs units, unless stated otherwise. Parenthesized numbers denote powers of 10. Basic model parameters: $F_r = 1.00$; $h = 0.08$ (symbols and definitions used here are identical to those used in Paper I) \label{tbl-1}} 
\begin{tabular}{lrr}
 & & \\
Parameter & AE Aqr WD & 1987A NS \\
\tableline
 & & \\ 
Mass, $M$ (solar masses)                       & 8.9$(-01)$ & 1.6$(+00)$ 
\\
Initial turn-over angle, $\chi_{ini}$ (arcdegrees)
                                               & 1.4$(+00)$ & 1.3$(+00)$ 
\\
Angular momentum, $L_{xx}$                     & 2.2$(+49)$ & 2.3$(+48)$ 
\\
Average surface poloidal field, $B_s$          & 2.9$(+06)$ & 3.0$(+13)$ 
\\
TOV timescale, $t_{TOV}$ (yr)                  & 9.0$(+06)$ & 1.6$(+01)$ 
\\
Secular timescale, $t_{SEC}$ (yr)              & 5.0$(+08)$ & 8.5$(+00)$
\\
Present TOV time, $t_{now}$ (yr)               & 4.6$(+06)$ & 8.5$(+00)$ 
\\
Present central period, $P_{now}$              & 3.31$(+01)$ & 2.14$(-03)$  
\\
Present turn-over angle, $\chi_{now}$ (arcdegrees)         
                                               & 7.2$(+01)$ & 6.3$(+01)$
\\
Average spin-down time rate due to turn-over, $\left< \dot{P} \right>_t$
                                               & 5.6$(-14)$ & 1.7$(-12)$ 
\\
Average power loss due to turn-over, $\left< \dot{T} \right>_t$ 
                                               & 4.9$(+32)$ & 4.1$(+41)$ 
\\
 & & \\
\tableline
\end{tabular}
\end{center}
\end{table}  

\acknowledgements{\textbf{ACKNOWLEDGMENTS}}
The research reported here was supported by the Research Committee
of the University of Patras (C. Carath\'eodory's Research Project
1998/1932). This research has made use of NASA's Astrophysics Data System.


\begin{thebibliography}{Geroyannis(2002)}
\bibitem[Geroyannis(2001)]{ger01} Geroyannis, V. S. 2001 astro-ph/0103080 (Paper I)
\bibitem[Geroyannis(2002)]{ger02} Geroyannis, V. S. 2002 \apjs, to appear in Vol. 141 No. 2 (Paper II)
\bibitem[Geroyannis \& Papasotiriou(2000)]{gpa00} Geroyannis, V. S., \& Papasotiriou, P. J. 2000 ApJ, 534, 359 
\bibitem[Geroyannis \& Papasotiriou(2002)]{gpa02} Geroyannis, V. S., \& Papasotiriou, P. J. 2002 astro-ph/0206476 (Paper III)
\bibitem[Geroyannis, Papasotiriou, \& Valiaka(2002)]{gpv02} Geroyannis, V. S., Papasotiriou, P. J., \& Valiaka, I. 2002 astro-ph/0207024 (Paper IV)
\bibitem[Mestel \& Takhar(1972)]{mta72} Mestel, L., \& Takhar, H. S. 1972 MNRAS, 156, 419 (MT72)
\end{thebibliography}
\end{document}